\documentclass[prd,showpacs,preprint,nofootinbib]{revtex4}
\usepackage{graphicx,color,amsmath}
\usepackage{bm}
\usepackage{slashed}

\begin{document}
\title{Instanton effects on the twist-three light-cone distribution
amplitudes of pion and light scalar mesons above 1GeV}
\author{ Jin Zhang$^1$, Zhu-Feng Zhang$^2$ and Hong-Ying Jin$^1$,}
\affiliation{$^1$Institute of Modern Physics, Zhejiang University,
Hangzhou, Zhejiang, China\\
$^2$Department of Physics, Ningbo University, Ningbo, Zhejiang,
China}

\begin{abstract}
By use of single instanton approximation the twist-3 light-cone
distribution amplitudes of pion as well as the p-wave mesons,
\emph{i.e.}, $f_{0}(1370)$, $K_{0}^{\ast}(1430)$ and $a_{0}(1450)$
are investigated within the framework of QCD moment sum rules with
inclusion of instanton effects based on the valence quark model.
Results show that there is much change on light-cone distribution
amplitudes by the isospin- and chirality-dependent instanton
contribution compared with the instanton-free ones. We find the
intanton involved twist-3 LCDAs are non-positive-definite within
some range of momentum fraction and there are rapid changes at two
ends of momentum fraction. To guarantee the convergence of moments
by the method in this work a low instanton density should be
adopted, for instance $n_{c}=\frac{1}{2}\rm{fm^{-4}}$ is welcomed.
Possible ingredients which might have impact on the results are
briefly discussed. These light-cone distribution amplitudes may be
helpful to analyze exclusive heavy flavored processes.
\end{abstract}


\maketitle
\newpage

\section{introduction}
The hadronic light-cone distribution amplitude(LCDA or light-cone
wave functions) which parameterize the non-perturbative effects play
an important role in understanding exclusive hard processes in
QCD\cite{Lepage, efremov, braun}. The LCDA is also one of the key
ingredients in QCD factorization approach\cite{beneke} for
describing exclusive hadronic B decays which are helpful to
ascertain the quark content of light scalars\cite{cheng1, cheng2,
cheng3} and other observables of phenomenological interest.
Therefore if there is more information on LCDA there will be more
complete study on heavy-flavored exclusive processes consequently
many important parameters in standard model. To extract LCDA the
nonperturbative method should be employed due to its nonperturbative
nature. The QCD sum rules\cite{shifman} have been proved to be very
successful one in obtaining useful information in hadronic
region\cite{smr}. While it was found later after its born in order
to produce reasonable results of lowest-lying states of pseudoscalar
channel the instanton effect\cite{belavin, hooft1, hooft2, callan,
dyak} should be taken into account in QCD sum
rules\cite{shuryaka}\footnote{For thorough review on instanton in
QCD one can refer to\cite{instrew}.}. Then the role of instanton in
QCD sum rules is extensively investigated\cite{smrinst, fork}.
Recently it was noticed that instanton may be helpful in lifting the
mass degeneration in light scalar mesons above 1GeV\cite{zhang}. The
calculation of LCDAs within the framework of QCD sum rules was
introduced in\cite{chernyaka} by studying the correlation function
of currents with derivatives in this way the desired Gegenbauer
coefficients can be expressed in terms of the moments derived from
QCD sum rules. Using this method Chernyak and Zhitnitsky(CZ) were
able to describe many experimental data available at that
time\cite{chernyakb}. Marrying with lattice simulation the first two
Gegenbauer coefficients of pion and kaon were studied\cite{latt}
under this method.

In analyzing B decays to p-wave mesons\cite{babar, belle}(more
complete review on B decays, see\cite{bdecay}) such as
$f_{0}(1370)$, $K_{0}^{\ast}(1430)$, $a_{0}(1450)$ the LCDAs are
important input. Having noticed the significant implication of
instanton in producing the realistic mass spectral of $f_{0}$,
$K_{0}^{\ast}$ and $a_{0}$ above 1GeV\cite{zhang}, in this paper we
will make some effort to investigate the instanton effects on the
twist-three LCDAs of these mesons by the CZ method. The work in this
paper can be regarded as partly pursuing of Ref. \cite{zhang} and
the quark content are assigned as follows
\begin{eqnarray}
f_{0}&=&\frac{1}{\sqrt{2}}(u\bar{u}+d\bar{d}),\nonumber\\
K_{0}^{\ast}&=&d\bar{s},\nonumber\\
a_{0}&=&\frac{1}{\sqrt{2}}(u\bar{u}-d\bar{d}),\label{quarkcont}
\end{eqnarray}
The LCDAs of pion, as a widely used quantity in exclusive processes,
naturally attach attentions based on various methods from instanton
model\cite{pilcda}. To this point we would like to stress that
instanton is crucial in reproducing reasonable results of pion and
its partners in $0^{-}$ channel by QCD sum rules\cite{shuryaka}.
Perturbatively the intuition is that due to the chirality suppress
there should be larger size of $s\bar{s}$ component in $\omega$ than
$\eta$. If it were the case $\omega$ would be close to $\phi$ but
the fact is that $\omega$ nearly degenerates with $\rho$ while there
is considerable mass gap between $\omega$ and $\phi$. The empirical
spectrum is that $\eta$ is nearly pure SU(3) octet while much
smaller $s\bar{s}$ component in $\omega$. Then there arise the
puzzle why there is much larger splitting between pion and $\eta$
than that of $\rho$ and $\omega$ which is difficult to explain
perturbatively. The puzzle can be solved if the instanton is
considered. Due to the definite chirality of fermion zero mode there
is direct instanton contribution to pseudoscalar channel but not in
the vector and tensor ones and additionally this instanton-induced
correction is isospin-dependent thus pion and $\eta$ are lighter
than $\rho$ and $\omega$ and there is larger mass gap between pion
and $\eta$ than $\rho$ and $\omega$. Hence pion provides a natural
laboratory to study instanton and it is expected there should be
some effects of instanton on its twist-3 LCDAs. Therefore it is
meaningful to analyze the twist-3 pion LCDAs by CZ method with
inclusion of instanton. In fact the isospin-dependence of instanton
effects is also important to reproduce a realistic mass gap between
$a_{0}$ and $f_{0}$ in $0^{+}$ channel above 1GeV by QCD sum rules
assuming small size $s\bar{s}$ mixing into $f_{0}$\cite{zhang}. The
quark content of pion is as usual
\begin{equation}
\pi^{0}=\frac{1}{\sqrt{2}}(u\bar{u}-d\bar{d}).\nonumber
\end{equation}
It is hoped these twist-3 LCDAs may shed some light on the exclusive
hadronic B decays. We would like to emphasize that in QCD language a
real hadron should be described by a set of Fock states which each
one has the same quantum number as the hadron. For example
\begin{equation}
|K_{0}\rangle=\psi^{K}_{d\bar{s}}|d\bar{s}\rangle+\psi^{K}_{d\bar{s}g}|d\bar{s}g\rangle
+\psi^{K}_{d\bar{s}q\bar{q}}|d\bar{s}q\bar{q}\rangle+...,\label{kdecom}
\end{equation}
It is no doubt there are also twist-3 LCDAs introduced by higher
Fock states but we will not consider them here. In other words we
deal with leading Fock states or the valence quarks of the hadrons.
Additionally there were some efforts in obtaining pion LCDAs from
holographic QCD methods\cite{agaev} and nonlocal condensates based
on QCD sum rules\cite{mikh} as well as some work on the shape of
pion LCDAs\cite{pionsh}. These work gives some insights of pion
LCDAs to us.

The outline of this paper is as follows: in Sec.
\makeatletter\@Roman{2} we present the sum rules with inclusion of
instanton contributions to obtain the LCDAs of $f_{0}$,
$K_{0}^{\ast}$, $a_{0}$ and pion. In Sec. \makeatletter\@Roman{3}
the numerical results and discussions will be presented, finally we
summarize our conclusions in Sec. \makeatletter\@Roman{4}. An
appendix is given to show the vanishing of instanton contribution to
tensor moment sum rules within the method.

\section{Basic formulas}

\subsection{moments and light-cone distribution amplitudes}
Firstly we define the decay constants of scalar meson S and the
pseudoscalar meson P
\begin{eqnarray}
\langle 0 |\bar{q}_{2}q_{1}|S(p)\rangle&=&m_{S}f_{S},\nonumber\\
\langle 0
|\bar{q}_{2}i\gamma_{5}q_{1}|P(p)\rangle&=&\frac{f_{P}m_{P}^{2}}{m_{1}+m_{2}},\nonumber
\end{eqnarray}
where $m_{S}$, $m_{P}$ $m_{1}$ and $m_{2}$ are the mass of scalar
meson, pseudoscalar meson, $q_{1}$ and $q_{2}$, respectively. The
twist-3 LCDAs $\phi_{S}^{s}(u)$ and $\phi_{S}^{\sigma}(u)$ for the
scalar meson $S$ with quark content $q_{1}\bar{q}_{2}$ are defined
as\cite{cheng2}
\begin{eqnarray}
\langle
0|\bar{q}_{2}(z_{2})q_{1}(z_{1})|S(p)\rangle&=&m_{S}f_{S}\int_{0}^{1}due^{i(up\cdot
z_{2}+\bar{u}p\cdot z_{1})}\phi_{S}^{s}(u),\label{sscalar}\\
\langle0|\bar{q}_{2}(z_{2})\sigma_{\mu\nu}q_{1}(z_{1})|S(p)\rangle&=&-m_{S}f_{S}(p_{\mu}
z_{\nu}-p_{\nu}z_{\mu})\int_{0}^{1}due^{i(up\cdot
z_{2}+\bar{u}p\cdot
z_{1})}\frac{\phi_{S}^{\sigma}(u)}{6}\label{stensor},
\end{eqnarray}
the two twist-3 two-particle LCDAs $\phi_{P}^{p}(u)$ and
$\phi_{P}^{\sigma}(u)$ of pseudoscalar meson are defined
as\cite{ball1}
\begin{eqnarray}
\langle0|\bar{q}_{2}(z_{2})i\gamma_{5}q_{1}(z_{1})|P(p)\rangle&=&\frac{f_{P}m_{P}^{2}}{m_{1}+m_{2}}
\int_{0}^{1}due^{i(up\cdot
z_{2}+\bar{u}p\cdot z_{1})}\phi_{P}^{p}(u), \label{ppseud}\\
\langle
0|\bar{q}_{2}(z_{2})\sigma_{\mu\nu}\gamma_{5}q_{1}(z_{1})|P(p)\rangle&=&-\frac{i}{3}
\frac{f_{P}m_{P}^{2}}{m_{1}+m_{2}}
(p_{\mu}z_{\nu}-p_{\nu}z_{\mu})\int_{0}^{1}due^{i(up\cdot
z_{2}+\bar{u}p\cdot z_{1})}\phi_{P}^{\sigma}(u),\label{ptensor}
\end{eqnarray}
where $u$ always refers to the momentum fraction carried by one
quark and $\bar{u}=1-u$ is another quark momentum fraction;
$z=z_{2}-z_{1}$. Noticing the gauge-invariant Wilson path-ordered
integral
\begin{equation}
[z_{2}, z_{1}]=P\exp
[ig\int_{z_{1}}^{z_{2}}d\sigma_{\mu}A^{\mu}(\sigma)].\nonumber
\end{equation}
has been suppressed. The normalization of these four twist-3 LCDAs
are
\begin{eqnarray}
\int_{0}^{1}du\phi_{S}^{s}(u)=\int_{0}^{1}du\phi_{S}^{\sigma}(u)=1,\nonumber\\
\int_{0}^{1}du\phi_{P}^{p}(u)=\int_{0}^{1}du\phi_{P}^{\sigma}(u)=1,
\end{eqnarray}
To proceed firstly we remind the reader that there is no pure
zero-mode contribution to tensor moment sum rules(see the appendix)
since instanton effect is chirality-dependent thus we do not
consider $\phi_{S}^{\sigma}(u)$ and $\phi_{P}^{\sigma}(u)$ here. For
the instanton-free tensor moment sum rules to calculate twist-3
LCDAs of p-wave mesons above 1GeV one can refer to Ref.\cite{lu}.
For simplicity in this paper we study the two $\phi_{S}^{s}(u)$ and
$\phi_{S}^{p}(u)$. In fact we can only concentrate on the scalar
moment sum rules because the pseudoscalar one can be deduced from
the scalar one by some appropriate substitutions, therefore in
following we deal with the scalar sum rules only.

Generally the twist-three LCDA $\phi_{S}^{s}(u)$ has the following
form
\begin{equation}
\phi_{S}^{s}(u,
\mu)=1+\sum_{n=1}^{\infty}a_{n}(\mu)C_{n}^{1/2}(2u-1),
\end{equation}
where $C_{n}^{1/2}(x)$ is Gegenbauer polynomials of order $1/2$, the
lowest ones are\cite{grad}
\begin{gather}
C_{0}^{1/2}(x)=1, \quad C_{1}^{1/2}(x)=x, \quad C_{2}^{1/2}(x)=\frac{1}{2}(3x^{2}-1),\nonumber\\
C_{3}^{1/2}(x)=\frac{5}{2}x^{3}-\frac{3}{2}x, \quad
C_{4}^{1/2}(x)=\frac{1}{8}(35x^{4}-30x^{2}+3),\label{gegen}
\end{gather}
and the orthogonality relation is
\begin{equation}
\int_{-1}^{1}C_{n}^{1/2}(x)C_{m}^{1/2}(x)dx=\frac{2}{2n+1}\delta_{nm}.\label{gegenorth}
\end{equation}
From Eq.(\ref{sscalar}) one can easily derive
\begin{equation}
\langle0|\bar{q}_{1}(0)(iz\cdot
\overleftrightarrow{D})^{n}q_{2}(0)|S(p)\rangle=m_{S}f_{S}(p\cdot
z)^{n}\langle\xi_{s}^{n}\rangle,\label{momentdef}
\end{equation}
where
\begin{eqnarray}
\overleftrightarrow{D}=\overrightarrow{D}_{\mu}-\overleftarrow{D}_{\mu},\quad
\overrightarrow{D}_{\mu}=\overrightarrow{\partial}_{\mu}-igA_{\mu}^{a}t^{a},\nonumber\\
\langle\xi^{n}_{s}\rangle=\int_{0}^{1}du(2u-1)^{n}\phi_{S}^{s}(u,\mu).\label{moment}
\end{eqnarray}
From the orthogonal relation Eq.(\ref{gegenorth}) the Gegenbauer
moments $a_{n}$ can be expressed in terms of
$\langle\xi^{n}\rangle$, for our purpose
\begin{equation}
a_{2}=\frac{5}{2}\big(3\langle\xi^{2}\rangle-1\big),\quad
a_{4}=\frac{9}{8}\big(35\langle\xi^{4}\rangle-30\langle\xi^{2}\rangle+3\big).\label{lcdacoeff}
\end{equation}
The next step is to calculate the so-called moments appearing in
Eq.(\ref{momentdef}), to this end we consider the following
two-point correlation function with derivatives
\begin{eqnarray}
I_{n0}(z, q)&=&i\int d^{4}xe^{iqx}\langle
0|TO_{n}(x)O^{\dagger}(0)|0\rangle\nonumber\\
&=&(z\cdot q)^{n}I^{\rm{OPE}}_{n0}(q^{2})\label{scalarcorr}
\end{eqnarray}
with
\begin{equation}
O_{n}(x)=\bar{q}_{1}(x)(iz\cdot \overleftrightarrow{D})^{n}q_{2}(x),
\quad O^{\dagger}(0)=\bar{q}_{2}(0)q_{1}(0),\nonumber
\end{equation}
The above correlation function can be expressed in terms of the
operator product expansion. Up to leading order of $\alpha_{s}$ and
dimension-six we get\footnote{Noticing the operator product
expansion is different from Ref. \cite{lu} on the mass-dependent
condensates terms, but there is little impact on the results since
these terms are greatly suppressed by the quark mass. We find the
operator product expansion in Eq. (\ref{scalarope}) presents a well
extremum behavior.}
\begin{eqnarray}
I_{n0}(z, q)&=&(z\cdot
q)^{n}\bigg[-\frac{3}{8\pi^{2}}\frac{1}{n+1}q^{2}\ln\frac{-q^{2}}{\mu^{2}}
+\frac{3+n}{24}\langle \frac{\alpha_{s}}{\pi}G^{2}\rangle\nonumber\\
&-&\frac{1}{q^{2}}\Big(\frac{n+1}{2}m_{1}+m_{2}\Big)\langle
\bar{q}_{1}q_{1}\rangle-\frac{1}{q^{2}}\Big(m_{1}+\frac{n+1}{2}m_{2}\Big)\langle
\bar{q}_{2}q_{2}\rangle\nonumber\\
&-&\frac{1}{2q^{4}}m_{2}\langle g_{s}\bar{q}_{1}\sigma
Gq_{1}\rangle-\frac{1}{2q^{4}}m_{1}\langle g_{s}\bar{q}_{2}\sigma
Gq_{2}\rangle\nonumber\\
&+&\frac{4\pi}{27}\frac{\alpha_{s}}{q^{4}}\Big(n^{2}+3n-4\Big)\Big(\langle
\bar{q}_{1}q_{1}\rangle^{2}+\langle\bar{q}_{2}q_{2}\rangle^{2}\Big)\nonumber\\
&-&\frac{48}{9}\frac{\alpha_{s}}{q^{4}}\langle
\bar{q}_{1}q_{1}\rangle\langle
\bar{q}_{2}q_{2}\rangle\bigg].\label{scalarope}
\end{eqnarray}
with $n$ is even thus only the scalar even moments exist. This is
the theoretical side of the correlation function from the
quark-gluon dynamics. On the other hand Eq.(\ref{scalarcorr}) can
also be derived phenomenologically based on the dispersion relation
\begin{equation}
I_{n0}(q^{2})=\frac{1}{\pi}\int_{0}^{\infty}ds\frac{{\rm{Im}}I^{ph}_{n0}(s)}{s-q^{2}}+\rm{subtr.\,
const.},
\end{equation}
The imaginary part ${\rm{Im}}I_{n0}^{ph}(s)$ is obtained by
inserting a complete quantum sets $\sum|n\rangle\langle n\rangle$
into Eq.(\ref{scalarcorr}) which reads
\begin{equation}
{\rm{Im}}I_{n0}^{ph}(q^{2})=\pi
m_{S}^{2}f_{S}^{2}\langle\xi^{n}\rangle\delta(q^{2}-m_{S}^{2})+
\frac{3}{8\pi^{2}}\frac{1}{n+1}\pi
q^{2}\theta(q^{2}-s_{0}),\label{phdis}
\end{equation}
By equating the theoretical and phenomenological sides of $I_{n0}(z,
q)$ we get the sum rules
\begin{equation}
I^{OPE}_{n0}(q^{2})=\frac{1}{\pi}\int_{0}^{\infty}ds\frac{{\rm{Im}}I^{ph}_{n0}(s)}{s-q^{2}}
+\rm{subtr. \,const.},\label{thpheq}
\end{equation}
Substituting Eq.(\ref{scalarope}) and Eq.(\ref{phdis}) into
Eq.(\ref{thpheq}), taking Borel transformation and subtracting the
continuum contributions we arrive the desired scalar moment sum
rules
\begin{eqnarray}
m_{S}^{2}f_{S}^{2}\langle\xi_{s}^{n}\rangle
\exp\big[-\frac{m_{S}^{2}}{M^{2}}\big]&=&\frac{3}{8\pi^{2}}\frac{1}{n+1}\int_{0}^{s_{0}}
ds\,s\exp\big[-\frac{s}{M^{2}}\big]+\frac{3+n}{24}\langle\frac{\alpha_{s}}{\pi}G^{2}\rangle\nonumber\\
&+&\Big(\frac{n+1}{2}m_{1}+m_{2}\Big)\langle\bar{q}_{1}q_{1}\rangle
+\Big(m_{1}+\frac{n+1}{2}m_{2}\Big)\langle\bar{q}_{2}q_{2}\rangle\nonumber\\
&-&\frac{1}{2M^{2}}m_{2}\langle g_{s}\bar{q}_{1}\sigma
Gq_{1}\rangle-\frac{1}{2M^{2}}m_{1}\langle g_{s}\bar{q}_{2}\sigma
Gq_{2}\rangle\nonumber\\
&+&\frac{4\pi}{27}\frac{\alpha_{s}}{M^{2}}\Big(n^{2}+3n-4\Big)\Big[\langle\bar{q}_{1}q_{1}\rangle^{2}
+\langle\bar{q}_{2}q_{2}\rangle^{2}\Big]\nonumber\\
&-&\frac{48\pi}{9}\frac{\alpha_{s}}{M^{2}}
\langle\bar{q}_{1}q_{1}\rangle\langle\bar{q}_{2}q_{2}\rangle.\label{sclarsr}
\end{eqnarray}
Up to now all our analysis still confines in the conventional QCD
sum rules, in the coming subsection the instanton will take part in
the game.

\subsection{Inclusion of the instanton effects in moment sum rules}

In this subsection our calculation are in \emph{four-dimension
Euclidean space} unless explicitly point out. Instanton is the
nontrivial solution of classical field equation in four-dimension
Euclidean gauge-field theories which is first discovered by Belavin
\emph{et al} \cite{belavin}. Subsequently 't Hooft\cite{hooft2}
derived the instanton with topological charge $Q=1$ in
four-dimension Euclidean space
\begin{eqnarray}
A^{a}_{\mu}(x)&=&\frac{2}{g}\eta_{a\mu\nu}
\frac{(x-x_{0})_{\nu}}{(x-x_{0})^{2}+\rho^{2}},\label{inst}\\
G^{a}_{\mu\nu}(x)&=&-\frac{4}{g}\eta_{a\mu\nu}
\frac{\rho^{2}}{\big[(x-x_{0})^{2}+\rho^{2}\big]^{2}},\label{instten}
\end{eqnarray}
where $\rho$ is instanton size, $\eta_{a\mu\nu}$ is the t'Hooft
$\eta$ symbol, $x_{0}$ is an any point in four-dimension Euclidean
space called instanton center. In this instanton background field
there is quark zero-mode which represent the tunneling effects, for
our purpose we write the quark zero-mode propagator explicitly in
\emph{regular gauge}\footnote{For the propagator in singular gauge
on can refer to\cite{facc}.}
\begin{eqnarray}
S^{zm}(x, y;
x_{0})&=&\frac{\rho^{2}}{8\pi^{2}m^{\ast}}\frac{1}{\big[(x-x_{0})^{2}+\rho^{2}\big]^{3/2}}
\frac{1}{\big[(y-x_{0})^{2}+\rho^{2}\big]^{3/2}}\nonumber\\
&\times&\Big[\gamma_{\mu}\gamma_{\nu}\frac{1}{2}(1-\gamma_{5})\Big]
\otimes[\tau_{\mu}^{+}\tau_{\nu}^{-}],\label{instpr}
\end{eqnarray}
where $m^{\ast}$ is the effective mass and
\begin{equation}
\tau^{\pm}_{\mu}=(\bm{\tau}, \pm i), \quad
\bm{\tau}=\bm{\sigma},\label{tau}
\end{equation}
with the useful relations
\begin{eqnarray}
\tau^{a}\tau^{b}&=&\delta^{ab}+i\varepsilon^{abc}\tau^{c},\nonumber\\
\tau^{+}_{\mu}\tau^{-}_{\nu}&=&\delta_{\mu\nu}+i\eta_{a\mu\nu}\tau^{a},\nonumber\\
\tau^{-}_{\mu}\tau^{+}_{\nu}&=&\delta_{\mu\nu}+i\bar{\eta}_{a\mu\nu}\tau^{a}.\label{newmatrix}
\end{eqnarray}
The instanton contribution to the scalar moment sum rules is
obtained by substituting the zero-mode propagator Eq.(\ref{instpr})
into correlation function Eq.(\ref{scalarcorr})
\begin{eqnarray}
\int d^{4}xe^{iQx}\langle
0|TO_{n}(x)O^{\dagger}(0)|0\rangle&=&\frac{8}{\pi^{4}\,m_{1}^{\ast}
m_{2}^{\ast}}\int d^{4}xe^{iQx}\int d\rho n(\rho)\rho^{4}\int
d^{4}x_{0}\frac{1}{(x_{0}^{2}+\rho^{2})^{3}}\nonumber\\
&\times&\frac{1}{\Big[(x-x_{0})^{2}+\rho^{2}\Big]^{3/2}}(iz\cdot
\overleftrightarrow{D}_{I})^{n}\frac{1}{\Big[(x-x_{0})^{2}+\rho^{2}\Big]^{3/2}},\label{zmconta}
\end{eqnarray}
with
\begin{equation}
O_{n}(x)=\bar{q}_{10}(x)(iz\cdot
\overleftrightarrow{D}_{I})^{n}q_{20}(x),\quad
O^{\dagger}(0)=\bar{q}_{20}(0)q_{10},\nonumber
\end{equation}
where the integration over collective coordinates of instanton is
explicit, the anti-instanton contribution as well as traces over
$\gamma$ and SU(2) matrix are completed implicitly. The instanton
density has the simple form proposed in\cite{shuryaka}
\begin{equation}
n(\rho)=n_{c}\delta(\rho-\rho_{c}),\label{instden}
\end{equation}
But now one should notice that the covariant derivative in
Eq.\eqref{zmconta} contains the instanton background field
$A^{(I)}_{\mu}$ to guarantee gauge invariance
\begin{equation}
D^{I}_{\mu}=\partial_{\mu}-igA^{(I)a}_{\mu}{t}^{a},\label{instcoder}
\end{equation}
where $t^{a}$ is the SU(2) generator $t^{a}=\sigma^{a}/2$ with the
normalization condition
\begin{equation}
{\rm{tr}}[t^{a}t^{b}]=\frac{1}{2}\,\delta^{ab},\label{normalization}
\end{equation}
Firstly we take a close look at the covariant derivative
\begin{eqnarray}
(iz\cdot\overleftrightarrow{D}_{I})^{n}&=&\Big[iz\cdot(\overrightarrow{D}_{I}-\overleftarrow{D}_{I})\Big]^{n}\nonumber\\
&=&\Big\{iz_{\mu}\Big[(\overrightarrow{\partial}_{\mu}-igA^{(I)a}_{\mu}{t}^{a})-(\overleftarrow{\partial}_{\mu}
+igA^{(I)a}_{\mu}{t}^{a})\Big]\Big\}^{n},\label{instderexpan}
\end{eqnarray}
We can sort the terms in expansion of Eq.\eqref{instderexpan} into
two kinds. The first one only contains the differential operator
which acts on zero-mode propagator from left and right, the second
one is the remaining parts which include all the instanton field
involved terms. In following we will elucidate the complete
contribution from these two kinds at $n=2$ and $n=4$ by explicit
calculation.

Contribution to the correlation function of the first kind is easy
to calculate such that we can derive a general formula for $n$ as
follows\footnote{Without confusion we take $\rho_{c}\rightarrow\rho$
for simplicity.}
\begin{eqnarray}
\Pi^{\text{first}}_{n}(Q^{2})&=&\int d^{4}xe^{iQx}\langle
0|TO_{n}(x)O^{\dagger}(0)|0\rangle\nonumber\\
&=&\frac{8\rho^{4}}{\pi^{4}\,m_{1}^{\ast}
m_{2}^{\ast}}\int d^{4}x_{0}e^{iQx_{0}}\frac{1}{(x_{0}^{2}+\rho^{2})^{3}}\times A(2i)^{n}\nonumber\\
&\times&\Big(-iz\cdot\frac{\partial}{\partial Q}\Big)^{n}\int
d^{4}xe^{iQx}\frac{1}{(x^{2}+\rho^{2})^{n+3}},\label{zmcontb}
\end{eqnarray}
where $A$ is constant
\begin{eqnarray}
A&=&\Big[1+(-1)^{n}\Big]\cdot\frac{3}{2}\cdot\frac{5}{2}\cdot...\cdot\frac{2n+1}{2}\nonumber\\
&+&\sum_{k=1}^{n-1}C_{n}^{k}(-1)^{n+k}\cdot\frac{3}{2}\cdot\frac{5}{2}\cdot...
\cdot\frac{2(n-k)+1}{2}\times\frac{3}{2}\cdot\frac{5}{2}\cdot...\cdot\frac{2k+1}{2},\nonumber
\end{eqnarray}
It is worthy to mention that in deriving Eq.(\ref{zmcontb}) from
Eq.(\ref{zmconta}) the light-cone constraint $z^{2}=0$ is crucial
otherwise there will be big masses. Now the remaining work is
trivial with the help of the following formulae\cite{shuryakb, grad}
\begin{eqnarray}
\int
d^{4}x\frac{e^{iQx}}{(x^{2}+\rho^{2})^{\nu}}&=&\frac{2\pi^{2}}{\Gamma(\nu)}\Big(\frac{Q\rho}{2}\Big)^{\nu-2}
\frac{K_{2-\nu}(Q\rho)}{\rho^{2\nu-4}},\nonumber\\
\Big(\frac{d}{zdz}\Big)^{m}\Big[z^{\nu}K_{\nu}(z)\Big]&=&(-)^{m}z^{\nu-m}K_{\nu-m}(z),\nonumber\\
K_{-\nu}(z)&=&K_{\nu}(z),\label{aidedfor}
\end{eqnarray}
where $Q^{2}=-q^{2}$ and $K_{\nu}(z)$ is the MacDonald function.
When the smoke clears, we get the desired results for the first kind
contribution
\begin{eqnarray}
\Pi^{\text{first}}_{n}(Q^{2})&=&\int d^{4}xe^{iQx}\langle
0|TO_{n}(x)O^{\dagger}(0)|0\rangle\nonumber\\
&=&\frac{n_{c}\rho^{2}}{\pi m_{1}^\ast
m_{2}^\ast}\frac{2^{n+1}\big[1+(-1)^{n}\big](1+n)}{\Gamma(n+3)}
\Big[\Gamma\Big(\frac{n+1}{2}\Big)\Big]^{2}\nonumber\\
&\times&(z\cdot Q)^{n}Q^{2}K_{1}^{2}(Q\rho),\label{zmcontc}
\end{eqnarray}
We find the non-vanishing contribution of the second kind for $n=2$
is
\begin{equation}
\Pi^{\text{second}}_{2}(Q^{2})=i^{2}\int d^{4}xe^{iQx}S_{10}(0, x;
x_{0})(-2igz\cdot A^{(I)})^{2}S_{20}(x, 0;
x_{0}),\label{twoinstcontr}
\end{equation}
Substituting the zero-mode propagator Eq.\eqref{instpr} and
instanton field Eq.\eqref{inst} at regular gauge into
Eq.\eqref{twoinstcontr}, for definite we write its explicit form as
follows
\begin{eqnarray}
\Pi^{\text{second}}_{2}(Q^{2})&=&-\frac{4g^{2}}{64\pi^{4}m^{\ast}_{1}m^{\ast}_{2}}\int
d^{4}x\,e^{iQ\cdot x}\int d\rho\,n(\rho)\rho^{4}\int
d^{4}x_{0}\frac{1}{(x_{0}^{2}+\rho^{2})^{3/2}}
\nonumber\\
&\times&\frac{1}{\big[(x-x_{0})^{2}+\rho^{2}\big]^{3/2}}\,\gamma_{\mu}\gamma_{\nu}\frac{1}{2}\,(1-\gamma_{5})
\tau^{+}_{\mu}\tau_{\nu}^{-}\nonumber\\
&\times&z_{\sigma}\frac{2}{g}\,\eta_{a\sigma\delta}\frac{(x-x_{0})_{\delta}}{(x-x_{0})^{2}+\rho^{2}}\,t^{a}
z_{\rho}\frac{2}{g}\,\eta_{b\rho\gamma}\frac{(x-x_{0})_{\gamma}}{(x-x_{0})^{2}+\rho^{2}}\,t^{b}\nonumber\\
&\times&\frac{1}{(x_{0}^{2}+\rho^{2})^{3/2}}
\frac{1}{\big[(x-x_{0})^{2}+\rho^{2}\big]^{3/2}}\,\gamma_{\alpha}\gamma_{\beta}\frac{1}{2}\,(1-\gamma_{5})
\tau^{+}_{\alpha}\tau_{\beta}^{-}\nonumber\\
&=&-\frac{n_{c}\,\rho_{c}^{4}}{64\pi^{4}m^{\ast}_{1}m^{\ast}_{2}}\int
d^{4}x\,e^{iQ\cdot x}\int
d^{4}x_{0}\frac{1}{(x_{0}^{2}+\rho_{c}^{2})^{3}}\frac{1}{\big[(x-x_{0})^{2}+\rho_{c}^{2}\big]^{5}}\nonumber\\
&\times&z_{\sigma}\,\eta_{a\sigma\delta}(x-x_{0})_{\delta}z_{\rho}
\,\eta_{b\rho\gamma}(x-x_{0})_{\gamma}\nonumber\\
&\times&{\rm{tr}}\big[\gamma_{\mu}\gamma_{\nu}\frac{1}{2}(1-\gamma_{5})
\gamma_{\alpha}\gamma_{\beta}\frac{1}{2}(1-\gamma_{5})\big]{\rm{tr}}
\big[\tau^{+}_{\mu}\tau^{-}_{\nu}\tau^{a}\tau^{b}\tau^{+}_{\alpha}\tau^{-}_{\beta}\big],\label{puretwoinsantonb}
\end{eqnarray}
After a lengthy calculation we obtain the contribution from
instanton field involved part for $n=2$ in expansion\footnote{For
simplicity we replace $\rho_{c}$ by $\rho$.}
\begin{equation}
\Pi^{\text{second}}_{2}(Q^{2})=\frac{n_{c}\rho^{2}}{6m^{\ast}_{1}m^{\ast}_{2}}(z\cdot
Q)^{2}Q^{2}K^{2}_{1}(Q\rho).\label{finalpuretwoinst}
\end{equation}
where the isospin dependence have been included. One can find that
this part is comparable to the first kind contribution for $n=2$.

For $n=4$ the situation is more complicated when we expand
Eq.\eqref{instderexpan} order by order since the differential
operator $\partial_{\mu}$ can also act on instanton field and give
non-vanishing contribution. In considering this effects on instanton
field we find the following terms for the second kind contribution
to correlator
\begin{eqnarray}
\Pi^{\text{second}}_{4}(Q^{2})&=&i^{4}\int
d^{4}xe^{iQx}\bigg\{S_{10}(0, x;
x_{0})4\Big[z\cdot(\overrightarrow{\partial}-\overleftarrow{\partial})\Big]^{2}S_{20}(x,
0; x_{0})\bigg\}\big(-2igz\cdot
A^{(I)}\big)^{2}\nonumber\\
&+&i^{4}\int d^{4}xe^{iQx}S_{10}(0, x;
x_{0})\bigg[4z_{\mu}z_{\nu}z_{\alpha}z_{\beta}A^{(I)}_{\alpha}\Big(\partial_{\mu}\partial_{\nu}A^{(I)}_{\beta}\Big)
(-2ig)^{2}\nonumber\\
&+&(-2igz\cdot A^{(I)})^{4}\bigg]S_{20}(x, 0;
x_{0}),\label{fourinstcontr}
\end{eqnarray}
where for brevity the SU(2) generator index is suppressed. Some
effort later we obtain
\begin{equation}
\Pi^{\text{second}}_{4}(Q^{2})=\frac{3n_{c}\rho^{2}}{20m^{\ast}_{1}m^{\ast}_{2}}(z\cdot
Q)^{4}Q^{2}K^{2}_{1}(Q\rho).\label{finalfourinst}
\end{equation}
Notice that we have included the anti-instanton effects both in
Eq.\eqref{finalpuretwoinst} and Eq.\eqref{finalfourinst}. Combining
Eq.\eqref{zmcontc}, Eq.\eqref{finalpuretwoinst} and
Eq.\eqref{finalfourinst} we get the complete zero-mode contribution
to correlation function
\begin{equation}
\Pi_{2}^{\text{zm}}(Q^{2})=\Pi_{2}^{\text{first}}(Q^{2})+\frac{n_{c}\rho^{2}}{6m^{\ast}_{1}m^{\ast}_{2}}
(z\cdot Q)^{2}Q^{2}K^{2}_{1}(Q\rho),\label{completeinsttwo}
\end{equation}
for $n=2$ and
\begin{equation}
\Pi_{4}^{\text{zm}}(Q^{2})=\Pi_{4}^{\text{first}}(Q^{2})+\frac{3n_{c}\rho^{2}}{20m^{\ast}_{1}m^{\ast}_{2}}
(z\cdot Q)^{4}Q^{2}K^{2}_{1}(Q\rho),\label{completeinstfour}
\end{equation}
for $n=4$, from the two equations above one can find the
contribution of two kinds is comparable thus we can not omit one of
them simply.

After dealing with the quark zero-mode in the instanton background
field, now we calculate non-zero mode contribution to the
correlation function. For $n=2$ this part is
\begin{equation}
\Pi^{\text{nzm}}_{2}(Q^{2})=i^{2}\int
d^{4}xe^{iQx}S^{\text{nzm}}_{1}(0, x)(-2igz\cdot
A^{(I)})^{2}S^{\text{nzm}}_{2}(x, 0),\label{twoinstnzm}
\end{equation}
where $S^{\text{nzm}}_{1}(0, x)$ and $S^{\text{nzm}}_{2}(x, 0)$ are
the non-zero-mode propagators. In instanton background field the
complete form of quark propagator consists of zero-mode and
non-zero-mode parts
\begin{equation}
S_{I}(x,y)=S_{I}^{\text{zm}}(x,y)+S_{I}^{\text{nzm}}(x,y),\label{completeprop}
\end{equation}
In previous paragraphs we have completed all contribution to
correlation function from zero-mode in single instanton
approximation. To obtain the non-zero-mode contribution we need
$S_{I}^{\text{nzm}}(x,y)$, while $S_{I}^{\text{nzm}}(x,y)$ is a
quite complicated object. It was shown\cite{plb549-93, prc69-065211,
prd67-113009} that it is reliable to take massless free propagator
approximation for $S_{I}^{\text{nzm}}(x,y)$ if the zero-mode
contribution to Green function is maximal. Fortunately in present
calculation this requirement can be met since there is direct
instaton contribution to the correlation function we considered thus
it is convenient to take massless free propagator approximation
\begin{equation}
S_{I}^{\text{nzm}}(x,y)=\frac{1}{2\pi^{2}}\frac{\slashed{x}-\slashed{y}}{(x-y)^{4}},\label{masslessnzm}
\end{equation}
for non-zero-mode. At this point it is necessary to stress that in
vector channel this approximation is no longer valid otherwise the
vector current is not always conserved\cite{prd17-1583}. In this
case the propagator of non-zero-mode is very involved one can refer
to Ref.\cite{instrew} for details.

Combining Eq.\eqref{twoinstnzm} and Eq.\eqref{masslessnzm} we arrive
\begin{equation}
\Pi^{\text{nzm}}_{2}(Q^{2})=\frac{8n_{c}}{\pi^{2}}\int
d^{4}xe^{iQx}\frac{1}{x^{6}}\int
d^{4}x_{0}\frac{\big[z\cdot(x-x_{0})\big]^{2}}{\big[(x-x_{0})^{2}+\rho^{2}\big]^{2}},\label{nzmcontr}
\end{equation}
According to translation invariance $(x-x_{0})\rightarrow u$ we can
factorize out the the integral with respect to instanton center
$x_{0}$ thus there are two separate integral given by nonzero mode
and instanton field. Since $z^{2}=0$ it is easy to see the instanton
integral is vanishing therefore there is no contribution of nonzero
mode to correlation function. In other word nonzero mode and
instanton field do not ``entangle'' each other. The physical meaning
is that in this case the instanton field does not transfer momentum
from one quark to another. Similar analysis also hold for $n=4$.

To be consistent with Eq.\eqref{sclarsr} we should reformulate the
total instanton induced contributions in term of dispersion
relation. For this purpose noticing the properties of MacDonald
function under analytical continuation are\cite{abram, fork}
\begin{equation}
K_{\nu}(z)=
\begin{cases}\frac{i\pi}{2}e^{i\pi\nu/2}H_{\nu}^{(1)}(ze^{i\pi/2})&-\pi<argz\leq\frac{\pi}{2}\\
-\frac{i\pi}{2}e^{-i\pi\nu/2}H_{\nu}^{(1)}(ze^{-i\pi/2})&\frac{\pi}{2}<argz\leq\pi\label{macd}
\end{cases}
\end{equation}
in above expression $H_{\nu}^{(1)}(z)$ is the Hankel function of the
first kind
\begin{equation}
H_{\nu}^{(1)}(z)=J_{\nu}(z)+iY_{\nu}(z).\label{hank}
\end{equation}
where $J_{\nu}(z)$ and $Y_{\nu}(z)$ are the Bessel functions and
Neumann functions, respectively.  The last step is improving
Eq.(\ref{zmcontc}) so that it is consistent with Eq.(\ref{sclarsr}).
To this end noticing the cut structure of the Hankel functions
Eq.(\ref{hank}) one can find
\begin{equation}
{\rm{Im}}K_{1}^{2}(-i\rho\sqrt{s})=\frac{\pi^{2}}{2}
J_{1}(\rho\sqrt{s})Y_{1}(\rho\sqrt{s})+\text{singular
term},\label{impmac}
\end{equation}
As usual in terms of the dispersion relation we get the final
results improved by the Borel transformation and continuum
contribution subtracted. To be definite we collect the whole results
as follows
\begin{eqnarray}
m_{S}^{2}f_{S}^{2}\langle\xi_{s}^{2}\rangle\exp\big[-\frac{m_{S}^{2}}{M^{2}}\big]
&=&\Pi^{OPE}_{2}(s_{0},M^{2})\nonumber\\
&+&\,(-1)^{I}\frac{\pi n_{c}\rho^{2}}{3m_{1}^\ast
m_{2}^\ast}\int_{0}^{s_{0}}dssJ_{1}(\rho\sqrt{s})Y_{1}(\rho\sqrt{s})\exp\big[-\frac{s}{M^{2}}\big],
\label{secondmomentfinal}
\end{eqnarray}
for $\langle\xi_{s}^{2}\rangle$ and
\begin{eqnarray}
m_{S}^{2}f_{S}^{2}\langle\xi_{s}^{4}\rangle\exp\big[-\frac{m_{S}^{2}}{M^{2}}\big]
&=&\Pi^{OPE}_{4}(s_{0},M^{2})\nonumber\\
&+&\,(-1)^{I}\frac{8\pi n_{c}\rho^{2}}{45m_{1}^\ast
m_{2}^\ast}\int_{0}^{s_{0}}dssJ_{1}(\rho\sqrt{s})Y_{1}(\rho\sqrt{s})
\exp\big[-\frac{s}{M^{2}}\big],\label{fourthmomentfinal}
\end{eqnarray}
for $\langle\xi_{s}^{4}\rangle$ where the isospin dependence has
been considered and $\Pi^{OPE}_{n}$ is the right hand side of
Eq.\eqref{sclarsr}. It is easy to derive the pseudoscalar moment sum
rules from the scalar ones if we consider the effects of
$i\gamma_{5}$ carefully in calculations
\begin{eqnarray}
\frac{m_{P}^{4}f_{P}^{2}\langle\xi_{p}^{2}\rangle}{(m_{1}+m_{2})^{2}}
\exp\big[-\frac{m_{P}^{2}}{M^{2}}\big]
&=&\Pi^{ps,\,OPE}_{2}(s_{0},M^{2}) \nonumber\\
&-&(-1)^{I}\frac{\pi n_{c}\rho^{2}}{3m_{1}^\ast
m_{2}^\ast}\int_{0}^{s_{0}}dssJ_{1}(\rho\sqrt{s})Y_{1}(\rho\sqrt{s})
\exp\big[-\frac{s}{M^{2}}\big],\label{pseudsecmomentfinal}
\end{eqnarray}
for $\langle\xi_{p}^{2}\rangle$ and
\begin{eqnarray}
\frac{m_{P}^{4}f_{P}^{2}\langle\xi_{p}^{4}\rangle}{(m_{1}+m_{2})^{2}}
\exp\big[-\frac{m_{P}^{2}}{M^{2}}\big]
&=&\Pi^{ps,\,OPE}_{4}(s_{0},M^{2})\nonumber\\
&-&(-1)^{I}\frac{8\pi n_{c}\rho^{2}}{45m_{1}^\ast
m_{2}^\ast}\int_{0}^{s_{0}}dssJ_{1}(\rho\sqrt{s})Y_{1}(\rho\sqrt{s})
\exp\big[-\frac{s}{M^{2}}\big],\label{pseudfourmomentfinal}
\end{eqnarray}
for $\langle\xi_{p}^{4}\rangle$ where $\Pi^{ps,\,OPE}_{n}$
represents the following instanton-free sum rules
\begin{eqnarray}
\Pi^{ps,\,OPE}_{n}(s_{0},M^{2})&=&\frac{3}{8\pi^{2}}\frac{1}{n+1}\int_{0}^{s_{0}}
ds\,s\exp\big[-\frac{s}{M^{2}}\big]+\frac{n-3}{24}\langle\frac{\alpha_{s}}{\pi}G^{2}\rangle\nonumber\\
&+&\Big(\frac{n+1}{2}m_{1}-m_{2}\Big)\langle\bar{q}_{1}q_{1}\rangle
+\Big(-m_{1}+\frac{n+1}{2}m_{2}\Big)\langle\bar{q}_{2}q_{2}\rangle\nonumber\\
&+&\frac{1}{2M^{2}}m_{2}\langle g_{s}\bar{q}_{1}\sigma
Gq_{1}\rangle+\frac{1}{2M^{2}}m_{1}\langle g_{s}\bar{q}_{2}\sigma
Gq_{2}\rangle\nonumber\\
&+&\frac{4\pi}{27}\frac{\alpha_{s}}{M^{2}}\Big(n^{2}+3n-4\Big)\Big[\langle\bar{q}_{1}q_{1}\rangle^{2}
+\langle\bar{q}_{2}q_{2}\rangle^{2}\Big]\nonumber\\
&+&\frac{48\pi}{9}\frac{\alpha_{s}}{M^{2}}
\langle\bar{q}_{1}q_{1}\rangle\langle\bar{q}_{2}q_{2}\rangle.\nonumber
\end{eqnarray}
Obviously the instanton contribution in pseudoscalar channel is
opposite to scalar one which reflects the chirality-dependence of
instanton effects.

Now all the formulae needed have been fixed. The parameters which
will be adopted in our numerical analysis are as
follows\cite{cheng2, Khod, pdg}
\begin{gather}
\alpha_{s}=0.517, \quad
\langle\frac{\alpha_{s}}{\pi}G^{2}\rangle=0.012\pm0.006\rm{GeV^{4}},
\nonumber\\
\langle\bar{u}u\rangle=\langle\bar{d}d\rangle=-(0.225\pm0.15)^{3}{\rm{GeV}^{3}},
\quad\langle\bar{s}s\rangle=(0.8\pm0.2)\langle\bar{u}u\rangle,
\nonumber\\
m_{u}=0.004{\rm{GeV}},\,m_{d}=0.006{\rm{GeV}}, \quad
m_{s}=0.12{\rm{GeV}},
\nonumber\\
\langle g_{s}\bar{u}\sigma Gu\rangle=\langle g_{s}\bar{d}\sigma
Gd\rangle=0.8{\rm{GeV^{2}}}\langle\bar{u}u\rangle,\quad\langle g_{s}
\bar{s}\sigma Gs\rangle=0.8\langle g_{s}\bar{u}\sigma
Gu\rangle.\label{condensate}
\end{gather}
We read the masses and decay constants of $f_{0}$, $K_{0}^{\ast}$
and $a_{0}$ from\cite{zhang}
\begin{eqnarray}
m_{f_{0}}=1380{\rm{MeV}},& f_{f_{0}}=375\rm{MeV},\nonumber\\
m_{K_{0}^{\ast}}=1450{\rm{MeV}},& f_{K_{0}^{\ast}}=370\rm{MeV},\nonumber\\
m_{a_{_{0}}}=1480{\rm{MeV}}, & f_{a_{0}}=370\rm{MeV}.\label{mass}
\end{eqnarray}
The mass and decay constant pion are\cite{pdg}
\begin{equation}
m_{\pi}=140{\rm{MeV}},\quad f_{\pi}=130\rm{MeV}.\label{decayconst}
\end{equation}
All the parameters in Eq.(\ref{condensate}) as well as
Eq.(\ref{mass}) and Eq.(\ref{decayconst}) are taken at
$\mu=1\rm{GeV}$.

The remaining important parameters are the instanton related ones.
For the effective masses of quarks and instanton size the following
values are work well\cite{facc,zhang}
\begin{eqnarray}
m_{u}^{\ast}=m_{d}^{\ast}=86{\rm{MeV}},\nonumber\\
m_{s}^{\ast}=114\pm28\rm{MeV},\nonumber\\
\rho=\frac{1}{3}\,{\rm{fm}}=\frac{1}{0.6}\rm{GeV^{-1}}.\nonumber
\end{eqnarray}
The last parameter is the instanton density which still needs
improvement. The original value $n_{c}=\frac{1}{2}\rm{fm^{-4}}$ is
used widely\cite{shuryaka, hfor, dorok}, while the lattice
calculation suggested $n_{c}\sim\rm{1fm^{-4}}$\cite{chu}. The work
of Cristoforetti \textit{et al}\cite{cris} based on the interacting
instanton liquid model shown even a larger one was needed, i.e.
$n_{c}=3\rm{fm^{-4}}$, to reproduce the nucleon mass and the
low-energy constants in chiral perturbation theory. For this reason
we will investigate the sensitivity of the moments consequently the
LCDAs for different instanton density.

\section{results and discussions}
Firstly we present the selection rule of the threshold and Borel
window. The continuum(and exited states) as well as dimension-six
condensates contribution should be controllable, as
usual\cite{shifman, akhod, PhysRevD.71.014029, lu} we demand that in
the instanton-free sum rules the continuum contribution the part in
the dispersive integral from $s_{0}$ to $\infty$ should be less than
30\% the total perturbative dispersion integration which sets an
upper limit to us, the dimension-six condensates be no more than
15\% which sets an lower limit to us. If there is extremum within
the Borel window selected, we take it as our calculated value
otherwise the mid-value within the window will be adopted. Then we
turn on instanton contribution under same threshold and Borel window
since in this way there will be well comparison for the two cases.
Of course one can analyze the moments by separate threshold and
Borel window when instanton effect turn on, however in general the
sum rule is sensitive to threshold and Borel window that the
instanton effect may be smeared by the change inducing by the new
threshold and Borel window. In following discussions all the Borel
windows satisfy the selection rule unless explicitly state.

\begin{figure}
\begin{center}
\includegraphics[scale=0.75]{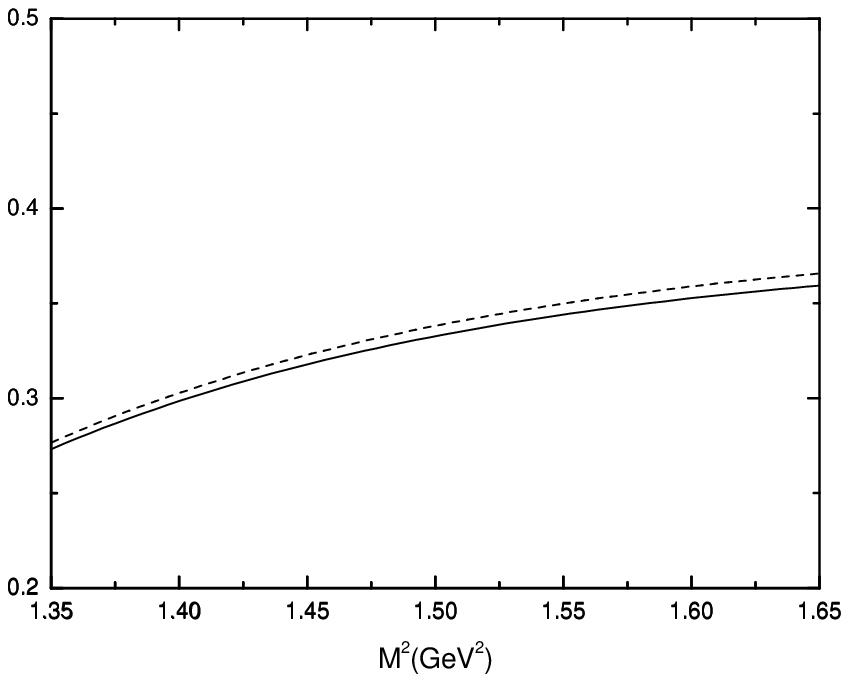}
\includegraphics[scale=0.75]{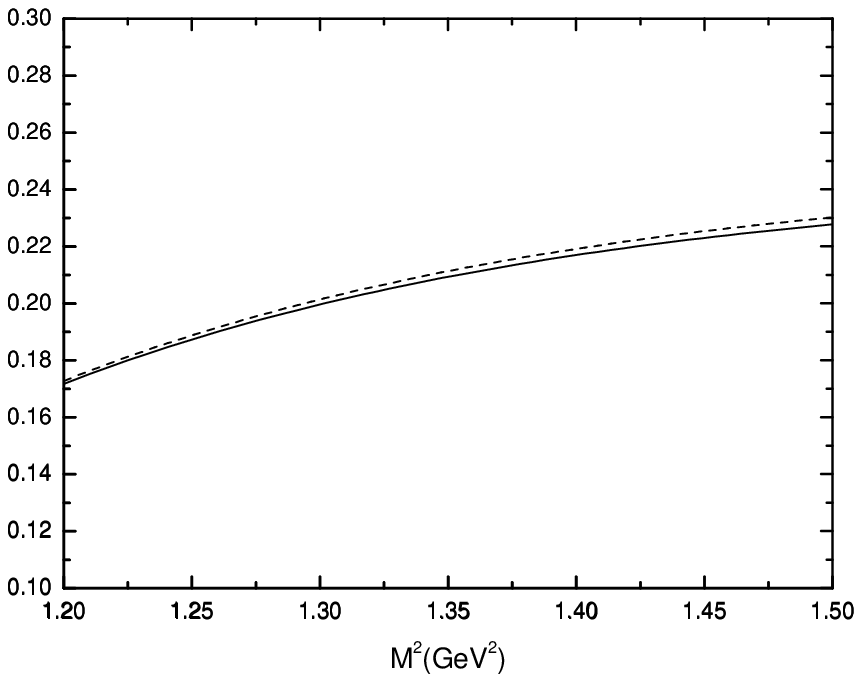}
\caption{Moments $\langle\xi_{p}^{2}\rangle$(left panel) and
$\langle\xi_{p}^{4}\rangle$(right panel) of pion from instanton-free
sum rules. Solid lines correspond to the central value of threshold
while the dashed lines represent the threshold increasing by
$0.1\rm{GeV^{2}}$ relative to central value(the same for
Fig.\,\ref{f0-moment}, Fig.\,\ref{k0-moment} and
Fig.\,\ref{a0-moment}). }\label{pi-moment}
\end{center}
\end{figure}

\begin{figure}
\begin{center}
\includegraphics[scale=0.75]{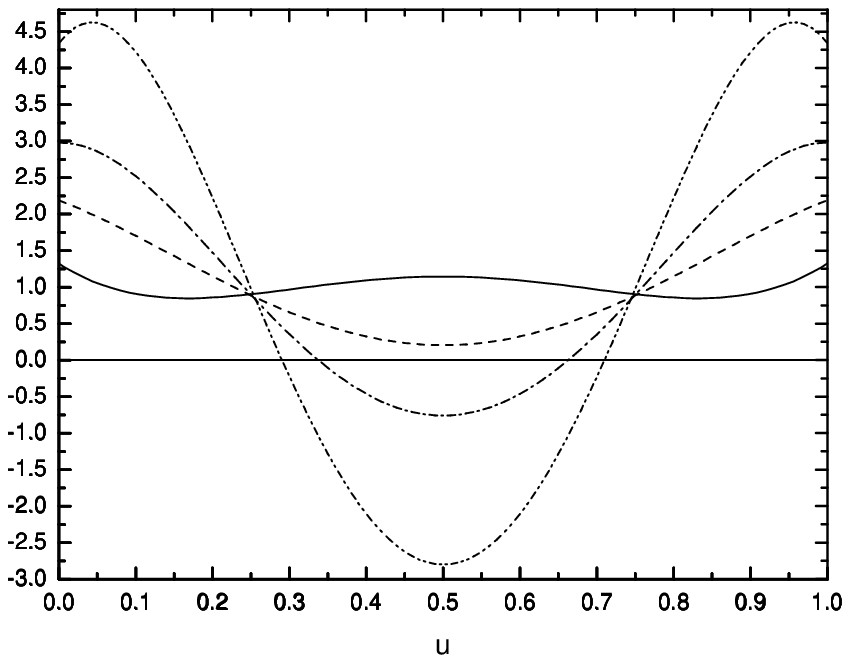}
\includegraphics[scale=0.75]{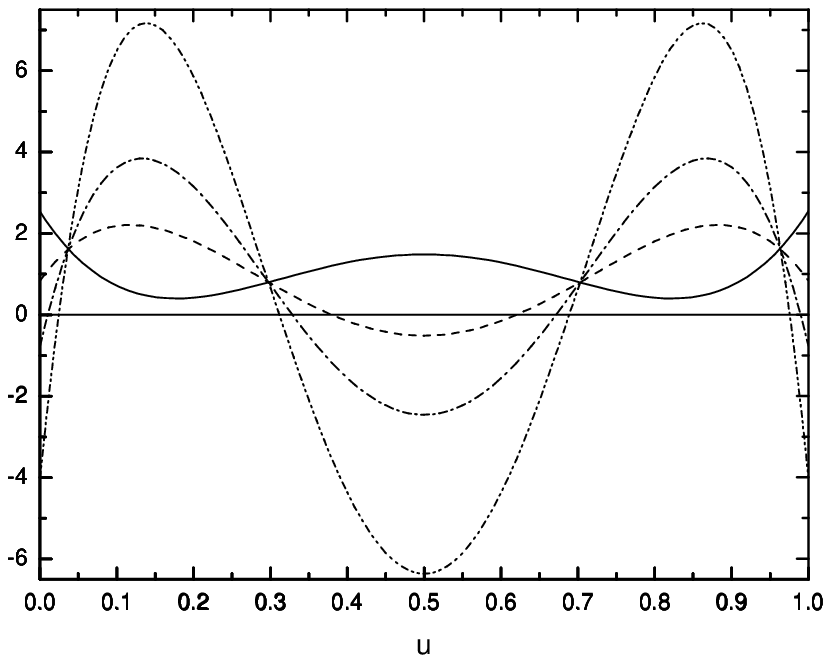}
\caption{Twist-3 light-cone distribution amplitudes of pion(left
panel) and $f_{0}(1370)$(right panel) as function of momentum
fraction $u$ at $\mu=1\rm{GeV}$ for different instanton density:
solid line $n_{c}=0$, dashed $n_{c}=\frac{1}{2}\rm{fm^{-4}}$ and
dashed-dot $n_{c}=1\rm{fm^{-4}}$, dash-dot-dot
$n_{c}=2\rm{fm^{-4}}$(the same for Fig.\,\ref{k0-a0-lcda}).
}\label{pi-f0-lcda}
\end{center}
\end{figure}

Along with the steps we find for pion the moment
$\langle\xi_{p}^{2}\rangle$ at threshold
$s_{0}=4.0\pm0.2\rm{GeV^{^{2}}}$ and Borel window $M^{2}\in[1.35,
1.65]\rm{GeV^{2}}$ as well as $\langle\xi_{p}^{4}\rangle$ at
$s_{0}=4.4\pm0.2\rm{GeV^{2}}$ and $M^{2}\in[1.2, 1.5]\rm{GeV^{2}}$
are stable from sum rules Eq.\eqref{pseudsecmomentfinal} and
Eq.\eqref{pseudfourmomentfinal} at $n_{c}=0$ respectively. The
mid-value are $\langle\xi_{p}^{2}\rangle=0.34$ and
$\langle\xi_{p}^{4}\rangle=0.21$ which are shown in
Fig.\,\ref{pi-moment}. When turning on the instanton effects we find
$\langle\xi_{p}^{2}\rangle=0.52$ and
$\langle\xi_{p}^{4}\rangle=0.36$ at $n_{c}=\frac{1}{2}\rm{fm^{-4}}$.
If increasing instanton density, for instance at
$n_{c}=1\rm{fm^{-4}}$ we find $\langle\xi_{p}^{2}\rangle=0.71$ and
$\langle\xi_{p}^{4}\rangle=0.50$. Obviously the moments increase
with instanton density increasing. Hence it is expected that at some
larger $n_{c}$ the second moment will be more than one, for instance
if we take $n_{c}=2\rm{fm^{-4}}$ as proposed in\cite{cris} we find
$\langle\xi_{p}^{2}\rangle=1.10$ and
$\langle\xi_{p}^{4}\rangle=0.80$, it is astonishing that the second
moment is more than 1! This result is unnatural since it is obvious
from Eq.(\ref{moment}) if assuming positive-definite LCDAs we have
\begin{eqnarray}
\langle\xi^{m}\rangle=\int_{0}^{1}du(2u-1)^{m}\phi(u,\mu)<
\langle\xi^{n}\rangle=\int_{0}^{1}du(2u-1)^{n}\phi(u,\mu)<1,\label{comp}
\end{eqnarray}
where
\begin{equation}
n, m=2,4,....., m>n.\nonumber
\end{equation}
But the LCDAs themselves are not measurable, to observe their exact
role we should convolute them with the hard scattering amplitudes
$T$ in exclusive processes. In fact the twist-2 LCDAs of
$f_{0}(980)$ given in\cite{cheng1} from CZ method, twist-3 ones of
pion\cite{ball1} as well as the work in Ref.\cite{prd56.1481,
prd71.054021, prd74.054023, prd81.074001, npb776.187} show
non-positive-definite behavior within some range of momentum
fraction. In considering this it seems that our results may indicate
non-positive-definite LCDAs. Indeed the instanton-involved LCDAs of
pion show this property as presented in Fig.\,\ref{pi-f0-lcda}
although the moments are still well convergent. In considering the
exact form of LCDAs and instanton density are not well known
nowadays so that the impact of high instanton density to LCDAs by
the CZ method seems to need further study. Our results show that it
seems that the sum rules Eq.\eqref{pseudsecmomentfinal} and
Eq.\eqref{pseudfourmomentfinal} do not allow too large instanton
density in order to get convergent moments.  The obtained LCADs of
pion for different instanton density is shown in the left panel of
Fig.\,\ref{pi-f0-lcda}.

\begin{figure}
\begin{center}
\includegraphics[scale=0.75]{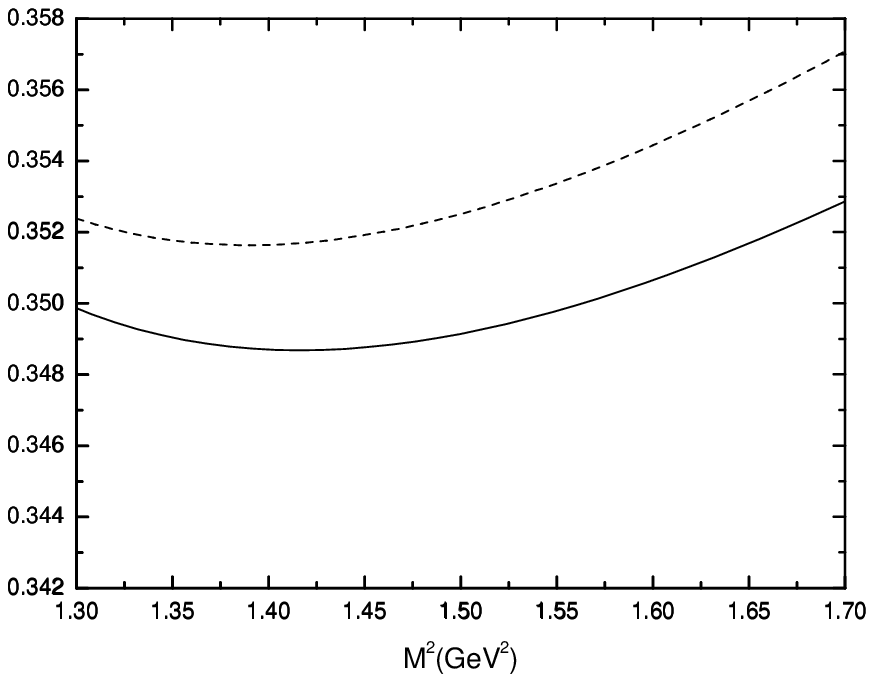}
\includegraphics[scale=0.75]{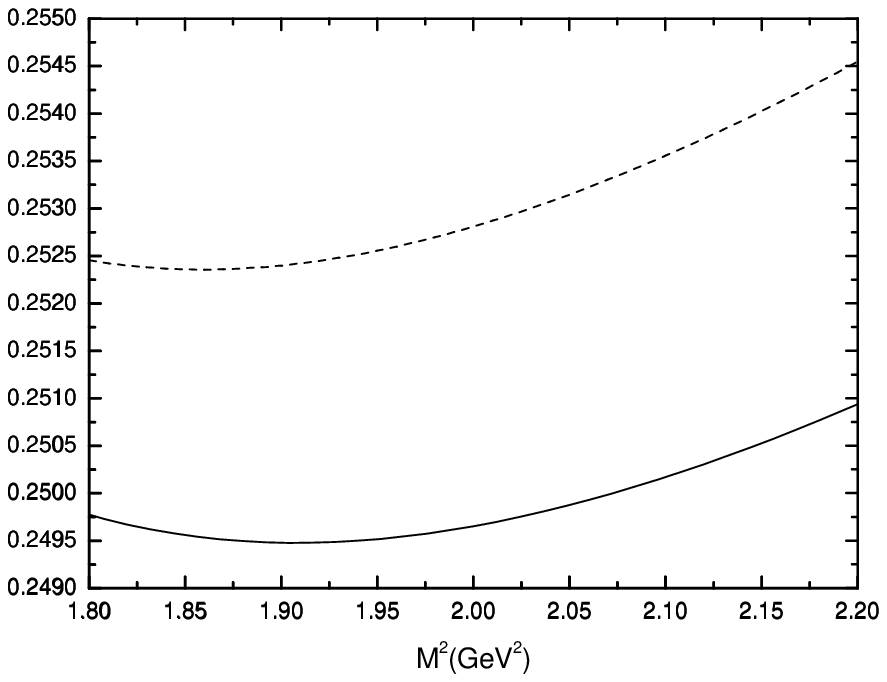}
\caption{Moments $\langle\xi_{s}^{2}\rangle$(left panel) and
$\langle\xi_{s}^{4}\rangle$(right panel) of $f_{0}(1370)$ at
$\mu=1\rm{GeV}$ from instanton-free sum rules. }\label{f0-moment}
\end{center}
\end{figure}

Then we turn to analyze $f_{0}(1370)$ with the quark content
assigned in Eq.(\ref{quarkcont}). The sum rules of this member is
very similar to the pion except some different condensates terms
induced by chirality. We can see the instanton contribution of
$f_{0}(1370)$ is same as pion since there are simultaneous changes
in its isospin and chirality relative to pion. At $n_{c}=0$ we get
the threshold and Borel window for $\langle\xi_{f_{0}}^{2}\rangle$
and $\langle\xi_{f_{0}}^{4}\rangle$ are
$s_{0}=4.7\pm0.2\rm{GeV^{2}}$, $M^{2}\in[1.3, 1.7]\rm{GeV^{2}}$ and
$s_{0}=4.8\pm0.2\rm{GeV^{2}}$, $M^{2}\in[1.8, 2.2]\rm{GeV^{2}}$,
respectively. Under the threshold and window there are well extremum
behavior both for $\langle\xi_{f_{0}}^{2}\rangle$ and
$\langle\xi_{f_{0}}^{4}\rangle$ which are shown in
Fig.\,\ref{f0-moment}. The extremum within the range of threshold
are very stable we obtain $\langle\xi_{f_{0}}^{2}\rangle=0.35$,
$\langle\xi_{f_{0}}^{4}\rangle=0.24$. When the instanton effects
turn on we find the mid-value are
$\langle\xi_{f_{0}}^{2}\rangle=0.55$ and
$\langle\xi_{f_{0}}^{4}\rangle=0.34$ for
$n_{c}=\frac{1}{2}\rm{fm^{-4}}$ as well as
$\langle\xi_{f_{0}}^{2}\rangle=0.76$ and
$\langle\xi_{f_{0}}^{4}\rangle=0.44$ for $n_{c}=1\rm{fm^{-4}}$.
Similar to the case of $\pi$ at $n_{c}=2\rm{fm^{-4}}$ we find
$\langle\xi_{f_{0}}^{2}\rangle=1.17$ which more than 1. The moments
also increase as the instanton density take a larger value which is
understandable since the instanton contributions to pion and
$f_{0}(1370)$ are equivalent due to the combined effects of isospin
and chirality. Thus we conclude the instanton contribution is
positive to the moment sum rules of pion and $f_{0}(1370)$. The
LCDAs of $f_{0}(1370)$ are plotted in Fig.\,\ref{pi-f0-lcda}.

It is easy to understand the similar impact of instanton on the
LCDAs of pion and $f_{0}(980)$ since the dominant parts, i.e.,
perturbative dispersive integral and the instanton contribution of
the moment sum rules for pion and $f_{0}(1370)$ are equivalent
except the chirality-dependent condensates. In fact this similarity
also reflects in the LCDAs which can be observed clearly from
Fig.\,\ref{pi-f0-lcda}. One more important thing is that the LCDAs
is positive-definite when there is no instanton effects while when
the instanton involved there is strong impact on the profile of
LCDAs. Due to the chirality-dependent parts at the two ends of
momentum fraction the LCDAs of $f_{0}(1370)$ change more rapidly
than pion.

\begin{figure}
\begin{center}
\includegraphics[scale=0.75]{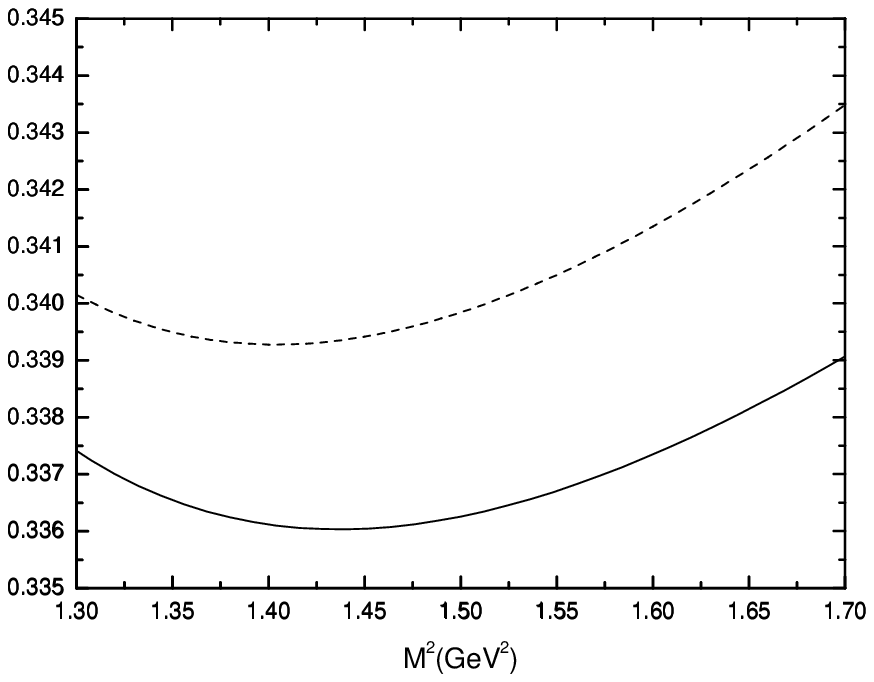}
\includegraphics[scale=0.75]{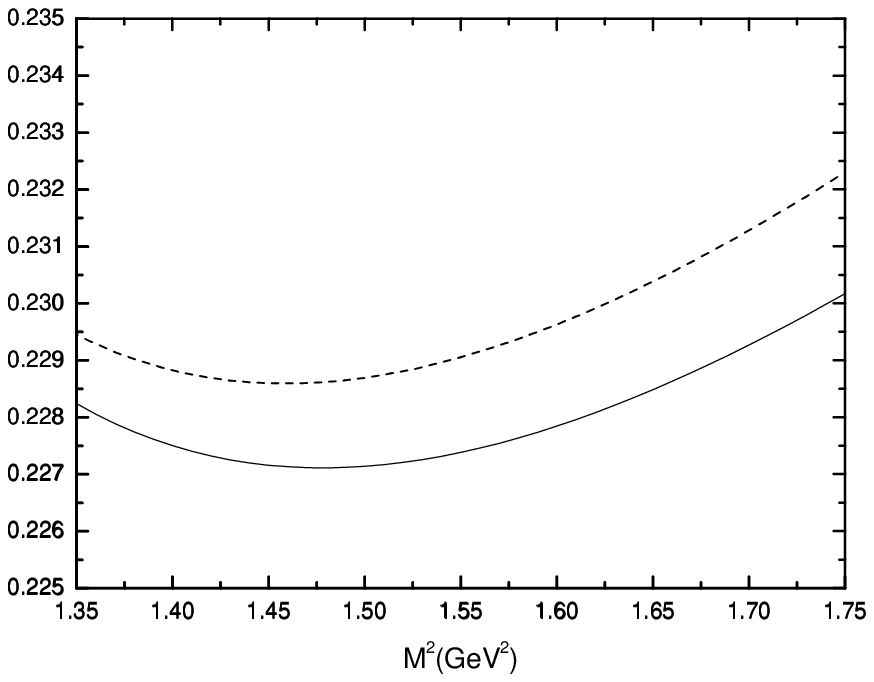}
\caption{Moments $\langle\xi_{s}^{2}\rangle$(left panel) and
$\langle\xi_{s}^{4}\rangle$(right panel) of $K_{0}^{\ast}(1430)$
from instanton-free sum rules. }\label{k0-moment}
\end{center}
\end{figure}

\begin{figure}
\begin{center}
\includegraphics[scale=0.75]{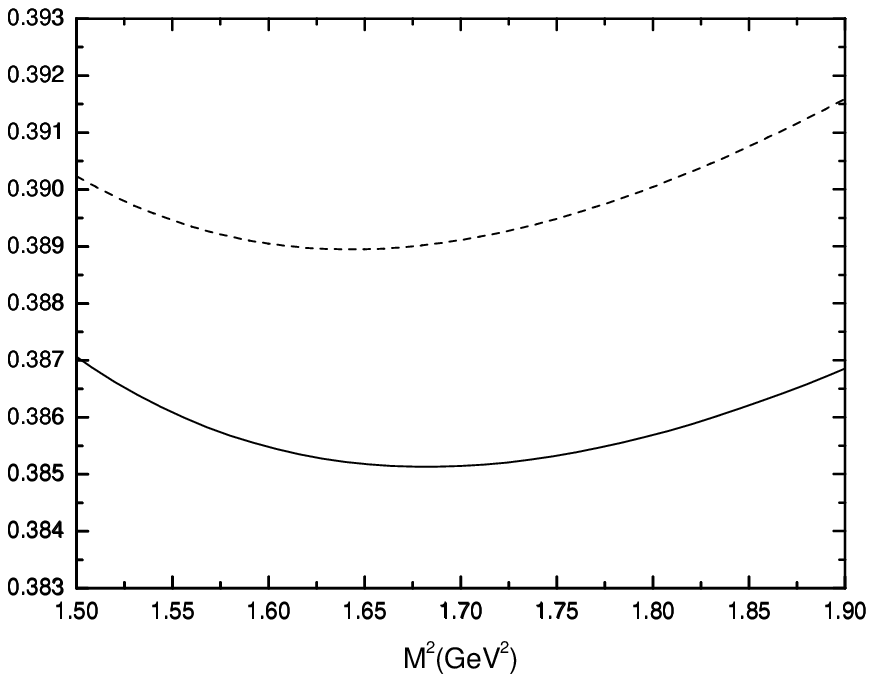}
\includegraphics[scale=0.75]{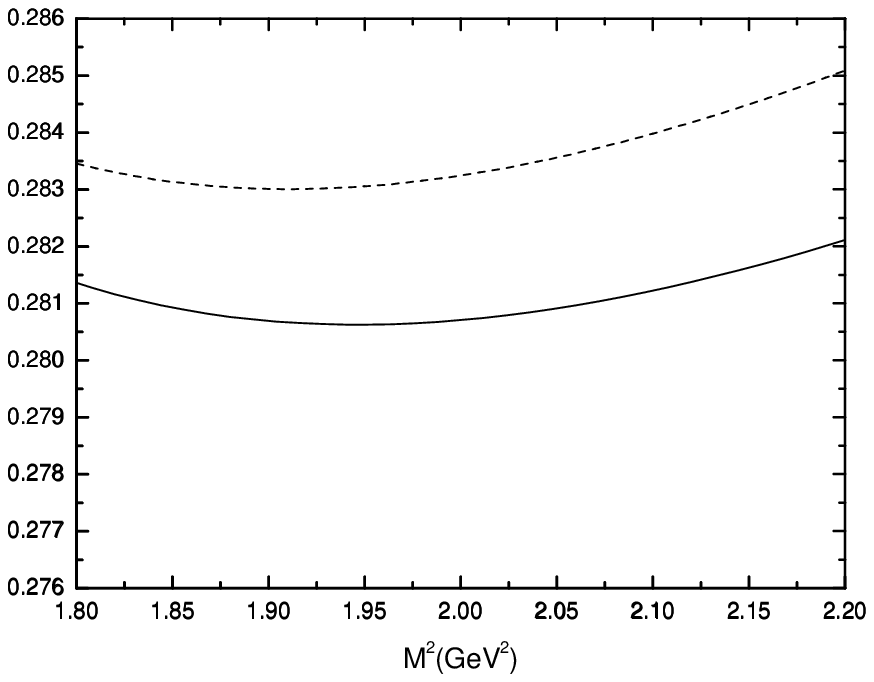}
\caption{Moments $\langle\xi_{s}^{2}\rangle$(left panel) and
$\langle\xi_{s}^{4}\rangle$(right panel) of $a_{0}(1450)$ from
instanton-free sum rules at $\mu=1\rm{GeV}$. }\label{a0-moment}
\end{center}
\end{figure}

\begin{figure}
\begin{center}
\includegraphics[scale=0.75]{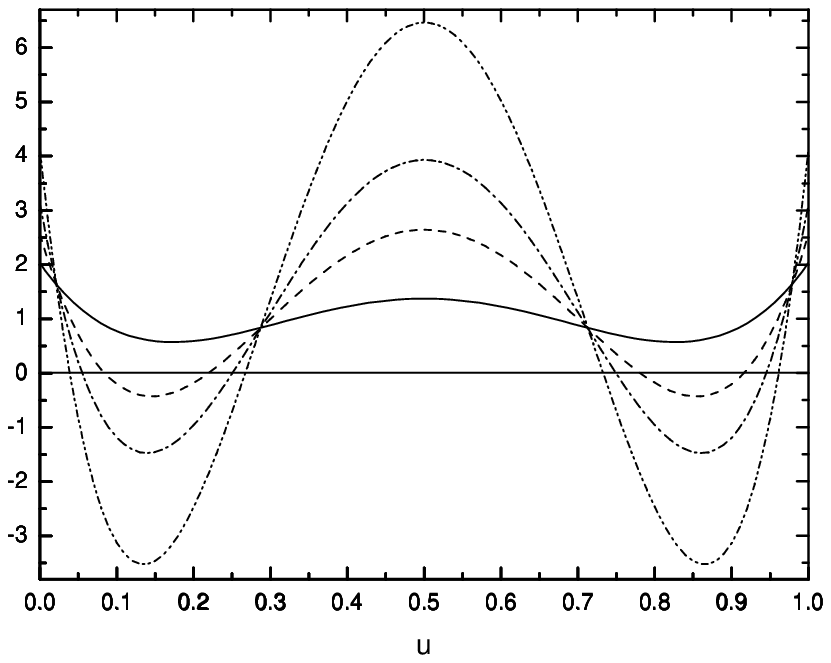}
\includegraphics[scale=0.75]{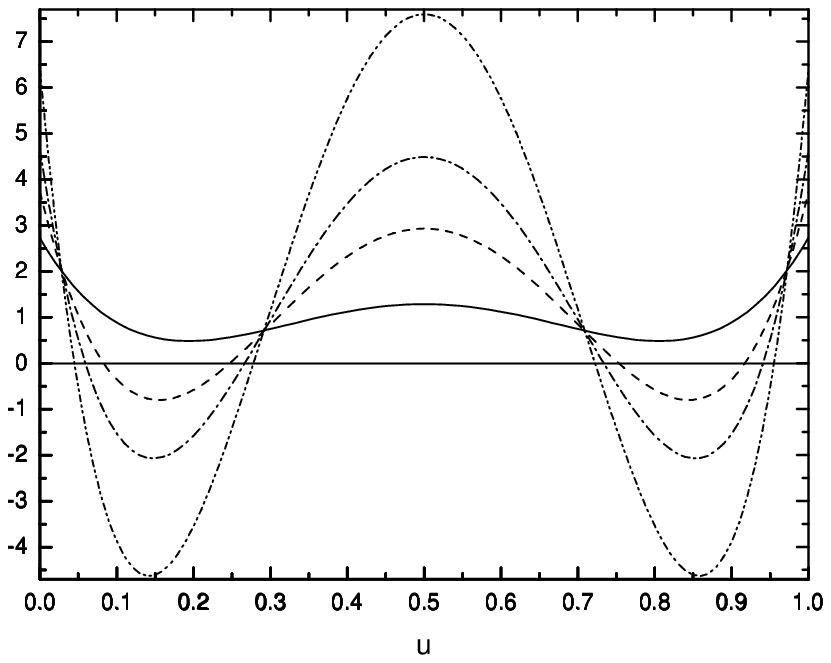}
\caption{Twist-3 light-cone distribution amplitudes of
$K_{0}^{\ast}(1430)$(left panel) and $a_{0}(1450)$(right panel) as
function of momentum fraction $u$ at $\mu=1\rm{GeV}$ for different
instanton density. }\label{k0-a0-lcda}
\end{center}
\end{figure}

The sum rules of $K_{0}^{\ast}(1430)$ and $a_{0}(1450)$ are nearly
the same since they share same isospin and chirality in addition to
the difference introduced by flavor symmetry breaking. The adopted
threshold and Borel window of $\langle\xi_{K_{0}^{\ast}}^{2}\rangle$
and $\langle\xi_{K_{0}^{\ast}}^{4}\rangle$ at $n_{c}=0$ are
$s_{0}=4.7\pm0.2\rm{GeV^{2}}$, $M^{2}\in[1.3, 1.7]\rm{GeV^{2}}$ and
$s_{0}=5.5\pm0.2\rm{GeV^{2}}$, $M^{2}\in[1.35, 1.75]\rm{GeV^{2}}$,
respectively. There is little change of the extremum corresponding
to different threshold both for the two moments, we get
$\langle\xi_{K_{0}^{\ast}}^{2}\rangle=0.34$ and
$\langle\xi_{K_{0}^{\ast}}^{4}\rangle=0.23$. When the instanton
effects involved the moments change lot even at low density
$n_{c}=\frac{1}{2}\rm{fm^{-4}}$. Both moments decrease compared with
the case $n_{c}=0$, we obtain
$\langle\xi_{K_{0}^{\ast}}^{2}\rangle=0.17$,
$\langle\xi_{K_{0}^{\ast}}^{4}\rangle=0.13$. If the instanton
density increases further, for instance $n_{c}=1\rm{fm^{-4}}$ we
find there is flipping of the second and fourth moments,
$\langle\xi_{K_{0}^{\ast}}^{2}\rangle=0.01$ and
$\langle\xi_{K_{0}^{\ast}}^{4}\rangle=0.04$ which is unsatisfactory
since it breaks the convergence. However at higher density
$n_{c}=2\rm{fm}^{-4}$ the convergence can recover while it develops
negative value --\,$\langle\xi_{K_{0}^{\ast}}^{2}\rangle=-0.32$ and
$\langle\xi_{K_{0}^{\ast}}^{4}\rangle=-0.16$. The LCDAs of
$K_{0}^{\ast}(1430)$ for different instanton density are plotted in
Fig.\,\ref{k0-a0-lcda}. It is clear that the profile of LCDAs of
$K_{0}^{\ast}(1430)$ with $n_{c}\neq0$ just reverse to pion and
$f_{0}(1370)$ which well indicates the conspiracy of chirality and
isospin dependence of instanton effects.

The case of $a_{0}(1450)$ runs in parallel to $K_{0}^{\ast}(1430)$.
The threshold and working window of $\langle\xi_{a_{0}}^{2}\rangle$
and $\langle\xi_{a_{0}}^{4}\rangle$ determined from with
instanton-free sum rules are $s_{0}=4.9\pm0.2\rm{GeV^{2}}$,
$M^{2}\in[1.5, 1.9]\rm{GeV^{2}}$ and $s_{0}=5.6\pm0.2\rm{GeV^{2}}$,
$M^{2}\in[1.8, 2.2]\rm{GeV^{2}}$, respectively. We obtain the
extremum $\langle\xi_{a_{0}}^{2}\rangle=0.39$ and
$\langle\xi_{a_{0}}^{4}\rangle=0.28$ within the threshold range.
After $n_{c}=\frac{1}{2}\rm{fm^{-4}}$ turning on both moments are
still well convergent and decrease to lower values:
$\langle\xi_{a_{0}}^{2}\rangle=0.20$,
$\langle\xi_{a_{0}}^{4}\rangle=0.17$. Similar to the case of
$\xi_{K_{0}^{\ast}}$, at $n_{c}=1\rm{fm^{-4}}$ we find there is also
flipping of the second and fourth
moments---\,$\langle\xi_{a_{0}}^{2}\rangle=0.01$ and
$\langle\xi_{a_{0}}^{4}\rangle=0.07$---\,which shows breakdown of
convergence of moment, at $n_{c}=2\rm{fm^{-4}}$ we can get
$\langle\xi_{a_{0}}^{2}\rangle=-0.37$ and
$\langle\xi_{a_{0}}^{4}\rangle=-0.13$ which shows negative value but
the convergence recovers. The moments and LCDAs of $a_{0}$ are
plotted in Fig.\,\ref{a0-moment} and right panel of
Fig.\,\ref{k0-a0-lcda}, respectively.

The instanton-involved twist-3 LCDAs calculated in this work present
nontrivial properties, to some extent is unexpected. Although in
principle non-positive-definite LCDAs are allowed, since here we
lack of direct test of these LCDAs in exclusive processes we would
like to give some tentative discussion on the possible ingredients
which are not mentioned above and may have some impacts on our
results.

\begin{itemize}
\item  \textbf{multi-instantons}
\end{itemize}
For simplicity we adopt single instanton approximation in our
calculation, in fact there might be multi-instanton contribution to
the correlation function. But the results in Ref.\cite{dorok}
indicate that in singular gauge multi-instanton contribution to the
pion correlator is coincident with that given by single-instanton
approximation. While physical results should be gauge-independent
thus we conjecture it might be reasonable to utilize
single-instanton approximation in our calculation. On the other hand
on can easily see it is very difficult to work out the instanton
contribution in singular gauge for $n\neq0$.

\begin{itemize}
\item  \textbf{instanton density}
\end{itemize}
It is obvious that the instanton density is crucial to obtain
convergent results. Low instanton density is welcomed in our
calculation. The density $\frac{1}{2}{\rm{fm}^{-4}}$ is widely used
under single instanton approximation which give many reasonable
results. In a way this indicates that low instanton density is
consistent with single approximation. In other word it seems that
high instanton density is questionable at single instanton
approximation. Maybe it is the main reason that at high density the
convergence is lost.

\begin{itemize}
\item  \textbf{subleading Fock states}
\end{itemize}
From Eq.(2) one can see we use valence model to investigate twist-3
LCDAs. The nonleading Fock states also may contribute twist-3
component via the mixing with other  wave functions.[30]. This
correction can be added by using the renormalization group, so it is
less relative to this work.

\section{conclusions}

In the present work we have investigated the instanton effect by
single instanton approximation on the twist-3 LCDAs of pion,
$f_{0}(1370)$, $K_{0}^{\ast}(1430)$ and $a_{0}(1450)$ from valence
quark model within the framework of QCD moment sum rules. Results
illustrate that the instanton-free twist-3 LCDAs are always
positive-definite while the instanton-involved LCDAs show some
nontrivial properties. We find that low instanton density is
consistent with the method adopted in this work. Possible
ingredients which might have impact on the results are briefly
discussed. Nonetheless we hope these LCDAs may be helpful to some
heavy flavored exclusive processes since we conjecture the instanton
density may play some role of a tuning parameter in deriving
experiment favored results.

\section{acknowledgements}
This work is partly supported by NNSFC under Project No. 10775117
and the Funda- mental Research Funs for the Central Universities.

\begin{appendix}
\section{Vanishing of the pure zero-mode contribution to tensor moment sum rules}
We still work in four-dimension Euclidean space. Considering the
following two-pint correlation function
\begin{eqnarray}
\int d^{4}xe^{iqx}\langle
0|TO_{n}(x)O^{\dagger}(0)|0\rangle&=&\frac{1}{256\pi^{4}\,m_{1}^{\ast}
m_{2}^{\ast}}\int d^{4}xe^{iqx}\int d\rho n(\rho)\rho^{4}\int
d^{4}x_{0}\frac{1}{(x_{0}^{2}+\rho^{2})^{3}}\nonumber\\
&&\times\frac{1}{\Big[(x-x_{0})^{2}+\rho^{2}\Big]^{3/2}}(iz\cdot
\overleftrightarrow{D}_{I})^{n+1}\frac{1}{\Big[(x-x_{0})^{2}+\rho^{2}\Big]^{3/2}}\nonumber\\
&&\times{\rm{tr}}\big[\gamma_{\mu}\gamma_{\nu}(1-\gamma_{5})\sigma_{\alpha\beta}
\gamma_{\sigma}\gamma_{\rho}(1-\gamma_{5})\big]\nonumber\\
&&\times{\rm{tr}}\big[\tau^{+}_{\mu}\tau^{-}_{\nu}\tau^{+}_{\sigma}\tau^{-}_{\rho}\big]\label{zmten}
\end{eqnarray}
with
\begin{equation}
O_{n}(x)=\bar{q}_{10}(x)\sigma_{\alpha\beta}(iz\cdot
\overleftrightarrow{D}_{I})^{n+1}q_{20}(x),\quad
O^{\dagger}(0)=\bar{q}_{20}(0)q_{10},\nonumber
\end{equation}
where the integrations over instanton collective coordinates are
explicitly. In fact it is enough to concentrate on the trace part in
Eq. (\ref{zmten})
\begin{eqnarray}
{\rm{tr}}\Big[\tau^{+}_{\mu}\tau^{-}_{\nu}\tau^{+}_{\sigma}\tau^{-}_{\rho}\Big]&=&
{\rm{tr}}\Big[(\delta_{\mu\nu}+i\eta_{a\mu\nu}\tau^{a})(\delta_{\sigma\rho}+i\eta_{b\sigma\rho}\tau^{b})\Big]\nonumber\\
&&=2\delta_{\mu\nu}\delta_{\sigma\rho}-2\eta_{a\mu\nu}\eta_{a\sigma\rho}\nonumber\\
&&=2(\delta_{\mu\nu}\delta_{\sigma\rho}-\delta_{\mu\sigma}\delta_{\nu\rho}+\delta_{\mu\rho}\delta_{\nu\sigma})
-2\varepsilon_{\mu\nu\sigma\rho}\label{tautr}
\end{eqnarray}
where the use of Eq. (\ref{tau}) and Eq. (\ref{newmatrix}) have been
made. Then combining above result with the trace over $\gamma$
matrix we have
\begin{eqnarray}
\Big[2(\delta_{\mu\nu}\delta_{\sigma\rho}-\delta_{\mu\sigma}\delta_{\nu\rho}+\delta_{\mu\rho}\delta_{\nu\sigma})
-2\varepsilon_{\mu\nu\sigma\rho}\Big]\times2{\rm{tr}}\Big[\gamma_{\mu}\gamma_{\nu}(1-\gamma_{5})\sigma_{\alpha\beta}
\gamma_{\sigma}\gamma_{\rho}\Big]\nonumber\\
=-2\varepsilon_{\mu\nu\sigma\rho}\times2{\rm{tr}}\Big[\gamma_{\mu}\gamma_{\nu}(1-\gamma_{5})\sigma_{\alpha\beta}
\gamma_{\sigma}\gamma_{\rho}\Big]\label{trmult}
\end{eqnarray}
while in four-dimension Euclidean space
\begin{equation}
\gamma_{5}=\frac{1}{4!}\varepsilon_{\mu\nu\sigma\rho}\gamma_{\mu}\gamma_{\nu}\gamma_{\sigma}\gamma_{\rho},\nonumber
\end{equation}
Now one can see the whole trace part in Eq. (\ref{zmten}) vanishes
thus there is no pure zero-mode contribution to tensor moment sum
rules consequently the LCDA.

\end{appendix}

\end{document}